\documentclass[aps,floatfix,showpacs,superscriptaddress,showkeys]{revtex4}

\usepackage{amssymb,amsmath,amsfonts,latexsym,graphicx,epsfig,here}
\usepackage{mkPhys}

\newcommand{\bea}{\begin{eqnarray}}
\newcommand{\ena}{\end{eqnarray}}

\newcommand{\be}{\begin{equation}}
\newcommand{\en}{\end{equation}}

\newcommand{\nn}{\nonumber\\}

\newcommand{\la}{\langle}
\newcommand{\ra}{\rangle}
\newcommand{\Bla}{\Big\langle}
\newcommand{\Bra}{\Big\rangle}

\newcommand{\Tr}{\mbox{\rm{tr}}}
      
\begin{document}

\hfill DSF-2012-4 (Napoli), MZ-TH/12-31 (Mainz)

\title{Light baryons and their electromagnetic interactions\\ 
in the covariant constituent quark model}

\author{Thomas Gutsche}
\affiliation{Institut f\"ur Theoretische Physik, Universit\"at T\"ubingen,\\
Kepler Center for Astro and Particle Physics,\\
Auf der Morgenstelle 14, D-72076, T\"ubingen, Germany}

\author{Mikhail A. Ivanov}
\affiliation{Bogoliubov Laboratory of Theoretical Physics, \\
Joint Institute for Nuclear Research, 141980 Dubna, Russia}

\author{J\"{u}rgen G. K\"{o}rner}
\affiliation{Institut f\"{u}r Physik, Johannes Gutenberg-Universit\"{a}t, \\
D-55099 Mainz, Germany}

\author{Valery E. Lyubovitskij
\footnote{On leave of absence
from Department of Physics, Tomsk State University,
634050 Tomsk, Russia}}
\affiliation{Institut f\"ur Theoretische Physik, Universit\"at T\"ubingen,\\
Kepler Center for Astro and Particle Physics,\\
Auf der Morgenstelle 14, D-72076, T\"ubingen, Germany}

\author{Pietro Santorelli}
\affiliation{Dipartimento di Scienze Fisiche, Universit\`a di Napoli
Federico II, Complesso Universitario di Monte S. Angelo,
Via Cintia, Edificio 6, 80126 Napoli, Italy, and
Istituto Nazionale di Fisica Nucleare, Sezione di Napoli}

%\today

\begin{abstract}
We extend the confined covariant constituent quark model that was previously 
developed by us for mesons to the baryon sector. In our numerical
calculation we use the same values for the constituent quark masses and the 
infrared cutoff as have been previously used in the meson sector.
In a first application we describe the static properties
of the proton and neutron, and the $\Lambda$-hyperon (magnetic moments and 
charge radii) and the behavior of the nucleon form factors at low momentum 
transfers. We discuss in some detail the conservation of gauge invariance of
the electromagnetic transition matrix elements in the presence of a nonlocal 
coupling of the baryons to the three constituent quark fields.  

\end{abstract}


\maketitle

\section{Introduction}

We use the confined covariant constituent quark model (for 
short: covariant quark model) as dynamical input to calculate the 
electromagnetic transition matrix elements between light $(u,d,s)$ baryons.
In the  covariant quark model the current--induced transitions between 
baryons are calculated from two--loop Feynman diagrams with free quark 
propagators in which the high energy behavior of the loop integrations is 
tempered by Gaussian vertex 
functions~\cite{Faessler:2001mr,Faessler:2008ix,Faessler:2009xn,Branz:2010pq}.
An attractive new feature has recently been added to the covariant quark model 
inasmuch as quark confinement has now been incorporated in an effective way, 
i.e. there are no quark thresholds  and thus no free quarks in the relevant 
Feynman  diagrams~\cite{Branz:2009cd,Ivanov:2011aa}. We emphasize that the
covariant quark model described here is a truly frame--independent
field theoretical quark model in contrast to other constituent quark models 
which are basically quantum mechanical with built--in relativistic elements. 
%One of the advantages of the covariant quark model is that it allows one to
%calculate the transition form factors in the full accessible range of 
%$q^{2}$--values. 

%The main feature of the  covariant quark model with infrared confinement
%is the possibility to 
In the covariant quark model we use the same values for the constituent quark 
masses and the infrared
cutoff for all hadrons (mesons and baryons) independent of the hadron masses.
We believe that the formulation of the confined covariant quark model 
constitutes a 
major advance both from the conceptual and the practical point of view.
While an unconfined quark model is valid only for hadrons 
with masses low enough to lie below the sum of the constituent quark masses 
the confined covariant quark model can be applied to all hadrons regardless
of their masses. The viability of the 
improved covariant quark model was demonstrated in a number of applications 
to mesonic transitions in~\cite{Branz:2009cd}. 
The form factors of the $B(B_s)\to P(V)$ transitions were evaluated
in~\cite{Ivanov:2011aa}, in a parameter-free way, 
in the full kinematical region of momentum transfer.
As an application, the widths of some nonleptonic $B_s$ decays were
also calculated.
This approach was successfully applied to 
a study of the tetraquark state X(3872) and its strong and radiative decays
(see, Refs.~\cite{Dubnicka:2010kz,Dubnicka:2011mm}).

In the present paper we formulate the covariant quark model
with infrared confinement for the baryon sector. By keeping the same values
for the constituent quark masses and the infrared cutoff as had been used in 
the meson sector we are able to reduce the number of free model 
parameters in the baryon sector to essentially the set of  
baryon size parameters. As a first application we describe the static 
properties of the nucleon and the $\Lambda$-baryon (magnetic moments and 
charge radii) and 
the behavior of the nucleon form factors at low momentum transfer. 
In a forthcoming publication~\cite{next}
we are planning to study the rare decays of the $\Lambda_b$-baryon into the
$\Lambda_{s}$ or the neutron. The present paper provides the necessary input
for such a calculation in as much we determine the properties of the light
baryons from their electromagnetic interactions. 

The paper is structured as follows.
In Sec.~II we review the basic notions of our dynamical approach --- 
{\it the covariant quark model for baryons}. 
We present the interaction Lagrangian describing the nonlocal coupling
of a proton to its constituents, discuss the choice of interpolating
currents and the vertex function,
recall the compositeness condition for bound-state hadrons and show how the 
confinement ansatz is implemented in the baryon sector. 
In Sec.~III we include the electromagnetic interactions of quarks and 
charged baryons in a manifestly gauge--invariant way,
derive the Lagrangian describing the nonlocal interaction
of the baryon, quark and electromagnetic fields.
% to the first order in the electromagnetic charge.
In Sec.~IV we present the loop integration techniques 
that allow one to calculate the nucleon mass function and its
derivative and the electromagnetic form factors of the nucleons. 
By analytically verifying the pertinent Ward and Ward--Takahashi identities 
we discuss in some detail how gauge invariance is maintained in the 
electromagnetic transitions. Sec.~IV also contains our
numerical results for the magnetic moments and form factors of the proton 
and neutron. We find that a particular superposition of 
vector and tensor interpolating currents gives satisfactory results for the 
nucleon static properties and form factors at low energies.
In Sec.~V we extend our approach to describe the static properties of
the $\Lambda_{s}$ hyperon. We summarize our findings in Sec.~VI. 

\section{The covariant quark model for baryons}

The basic ingredients of the covariant quark model for baryonic three quark
states prior to the implementation of confinement can be found 
in~\cite{Faessler:2001mr,Faessler:2008ix,Faessler:2009xn,Branz:2010pq}. 
This includes a description of the structure of the Gaussian vertex factor, 
the choice of interpolating baryon currents as well as the compositeness 
condition for baryons. 
   
The new features introduced to the meson sector 
in~\cite{Branz:2009cd,Ivanov:2011aa} and applied to the baryon sector in 
this paper are both technical and conceptual. Instead of using Feynman
parameters for the evaluation of the two--loop baryonic quark model
Feynman diagram we now use Schwinger parameters. The technical advantage
is that this leads to a simplification of the tensor loop integrations 
inasmuch as the loop momenta occurring in the quark propagators can be 
written as derivative operators. Furthermore, the use of Schwinger parameters 
allows one to incorporate quark confinement in an effective way.
Details of these two new features of the covariant quark model have been 
described in~\cite{Branz:2009cd,Ivanov:2011aa}. 

Let us enumerate the number of model parameters that are needed in our
approach for the description of baryons. As stated in the introduction the 
values of the constituent quark masses and the universal confinement 
parameter are taken over from the meson sector. The coupling strength of a 
baryon to its constituent quarks is fixed by the compositeness condition. 
This leaves one with one size parameter for each baryon. For the
present paper we need the size parameters of the proton, neutron and 
$\Lambda_{s}$. Naturally we
use the same size parameter for the proton and the neutron. The size 
parameters are determined by a fit to the static e.m. properties and form 
factors of the nucleons and the $\Lambda_{s}$. We have added one more 
parameter to the set of basic parameters which 
describes the mixing between vector and tensor interpolating currents.  
%----------------------------------------------------------------------------
\subsection{Lagrangian and three-quark currents}
%----------------------------------------------------------------------------

Let us begin our discussion with the proton.
The coupling of a proton to its constituent quarks is described by 
the Lagrangian 
\be\label{eq:Lagr_str}
{\cal L}^{\,\rm p}_{\rm int}(x) = g_N \,\bar p(x)\cdot J_p(x) 
                           + g_N \,\bar J_p(x)\cdot p(x)\,,  
\en
where we make use of the same interpolating three-quark current 
$J_p(\bar J_p)$ as in Ref.~\cite{Ivanov:1996pz}
\bea
J_p(x) &=& \int\!\! dx_1 \!\! \int\!\! dx_2 \!\! \int\!\! dx_3 \, 
F_N(x;x_1,x_2,x_3) \, J^{(p)}_{3q}(x_1,x_2,x_3)\,, \nonumber\\
J^{(p)}_{3q}(x_1,x_2,x_3) &=& 
\Gamma^A\gamma^5 \, d^{a_1}(x_1) 
\cdot[\epsilon^{a_1a_2a_3} \, u^{a_2}(x_2) \,C \, \Gamma_A \, u^{a_3}(x_3)]\,,
\nn
\label{eq:current}\\ 
\bar J_p(x) &=& \int\!\! dx_1 \!\! \int\!\! dx_2 \!\! \int\!\! dx_3 \, 
F_N(x;x_1,x_2,x_3) \, \bar J^{(p)}_{3q}(x_1,x_2,x_3)\,,
\nn
\bar J^{(p)}_{3q}(x_1,x_2,x_3) &=& 
[\epsilon^{a_1a_2a_3} \, 
\bar u^{a_3}(x_3)\, \Gamma_A \,C\, \bar u^{a_2}(x_2)]  
\cdot \bar d^{a_1}(x_1) \gamma^5\Gamma^A\,.
\nonumber
\ena
The matrix $C=\gamma^{0}\gamma^{2}$ is
the usual charge conjugation matrix and the $a_i$ $(i=1,2,3)$ are color 
indices. There are two possible kinds of nonderivative three-quark currents:
$\Gamma^A\otimes \Gamma_A = \gamma^\alpha\otimes \gamma_\alpha$~(vector
current) and 
$\Gamma^A\otimes \Gamma_A = \frac{1}{2} \, \sigma^{\alpha\beta}\otimes 
\sigma_{\alpha\beta}$~(tensor current) with  
$\sigma^{\alpha\beta}
= \tfrac{i}{2}(\gamma^\alpha\gamma^\beta-\gamma^\beta\gamma^\alpha).$ 
The interpolating current of the neutron and the corresponding Lagrangian are 
obtained from the proton case via $p \to n$ and 
$u \leftrightarrow d$. As will become apparent later on, one has to consider
a general linear superposition of the vector and tensor currents according to 
\be
\label{eq:superpo}
J_N = x J^T_N + (1-x) J^V_N\,, \quad N = p, n 
\en
where the mixing parameter $x$ extends from zero to one \,($0\le x \le 1$). 
When taking the nonrelativistic limit of the vector and tensor currents one 
finds that the two currents become degenerate. The limiting currents for the
proton and the neutron read  
\be 
J^T_p \equiv J^V_p \ = \ 
\epsilon^{a_1a_2a_3} \ \vec{\sigma} 
\, \psi_{d}^{a_1} \ ( \psi_{u}^{a_2} \, \sigma_2 \vec{\sigma} 
\, \psi_{u}^{a_3} )\,, 
\qquad
J^T_n \equiv J^V_n  \ = \ 
\epsilon^{a_1a_2a_3} \ \vec{\sigma} 
\, \psi_{u}^{a_1} \ ( \psi_{d}^{a_2} \, \sigma_2 \vec{\sigma} 
\, \psi_{d}^{a_3} ) 
\,,
\label{eq:NR-limit} 
\en 
where $\psi_{u,d}$ are the upper components of the respective Dirac quark 
spinor fields and where $\sigma_i$ are Pauli spin matrices.

Most of the properties of the nucleons are only weakly dependent on the 
choice of interpolating currents. However, in order to get the correct
value for the charge radius of the neutron, one needs to use the superposition 
of currents Eq.~(\ref{eq:superpo}) even though the currents $J^{T}$ and 
$J^{V}$ become degenerate in the nonrelativistic limit. In view of the fact
that a nonrelativistic description of the neutron gives zero values for the 
charge radius of the neutron \cite{deAraujo:2003ke} this is an indication that 
relativistic corrections play a crucial role for the desciption of the 
neutron charge radius.

The vertex function $F_N$ characterizes the finite size 
of the nucleon. We assume that the vertex function is real and 
the same for the proton and the neutron. To satisfy translational invariance 
the function $F_N$ has to satisfy the identity 
\be
F_N(x+a;x_1+a,x_2+a,x_3+a) \, = \, 
F_N(x;x_1,x_2,x_3) 
\label{eq:trans_inv}
\en
for any given 4-vector $a\,$. 
We use the following representation for the vertex function 
\be
F_N(x;x_1,x_2,x_3) \, = \, \delta^{(4)}(x - \sum\limits_{i=1}^3 w_i x_i) \;  
\Phi_N\biggl(\sum_{i<j}( x_i - x_j )^2 \biggr)\,,
\label{eq:vertex}
\en 
where $\Phi_N$ is the correlation function of the three constituent 
quarks with the coordinates $x_1$, $x_2$, $x_3$ and masses $m_1$, $m_2$, $m_3$,
respectively. The variable $w_i$ is defined by 
$w_i=m_i/(m_1+m_2+m_3)$ such that $\sum_{i=1}^3 w_i=1$. Note that 
$F_N(x;x_1,x_2,x_3)$ is symmetric in the coordinates $x_{i}$, i.e. symmetric
under $x_{i} \leftrightarrow{x_{j}}$. 

We shall make use of the Jacobi coordinates $\rho_{1,2}$ and the CM 
coordinate $x$ which are defined by

\bea
x_1 & = & x \, + \, \tfrac{1}{\sqrt{2}}\,w_3\, \rho_1 
            \, - \, \tfrac{1}{\sqrt{6}} \, (2w_2 + w_3)\,\rho_2 \,,
\nn
x_2 & = & x \, + \, \tfrac{1}{\sqrt{2}}\,w_3\, \rho_1
            \, + \, \tfrac{1}{\sqrt{6}}\, (2w_1 + w_3)\,\rho_2\,, 
\nn
x_3 & = & x \, - \, \tfrac{1}{\sqrt{2}}\,(w_1+w_2)\,\rho_1
            \, + \, \tfrac{1}{\sqrt{6}}\,(w_1 - w_2)\,\rho_2\,.
\label{eq:Jacobi}
\ena 
The CM coordinate is given by $x \,= \, \sum_{i=1}^3 w_i x_i$. 
In terms of the Jacobi coordinates one obtains
\be
\sum\limits_{i<j}( x_i - x_j )^2 = \rho_1^2 + \rho_2^2\,.
\label{eq:relative} 
\en 
Note that the choice of Jacobi coordinates is not unique.
With the above choice Eq.~(\ref{eq:Jacobi}) one readily arrives at the 
following representation for the
correlation function $\Phi_N$ in Eq.~(\ref{eq:vertex})  
\bea
\Phi_N\biggl(\sum_{i<j}( x_i - x_j )^2 \biggr) &=&
\int\!\frac{d^4p_1}{(2\pi)^4}\!\int\!\frac{d^4p_2}{(2\pi)^4}\,
e^{-ip_1(x_1-x_3) - ip_2(x_2-x_3)}\,\bar\Phi_N(-P^2_1-P^2_2)\,,
\label{eq:Fourier}\\
\bar\Phi_N(-P^2_1-P^2_2) &=&
\tfrac{1}{9}\int\!d^4\rho_1\int\!d^4\rho_2
\,e^{iP_1\rho_1 + iP_2\rho_2}\,\Phi_N(\rho^2_1+\rho^2_2)\,,
\nn
&&P_1=\tfrac{1}{\sqrt{2}}(p_1+p_2)\,,\qquad 
  P_2=-\tfrac{1}{\sqrt{6}}(p_1-p_2)\,.
\nonumber
\ena 
Even if the above choice of Jacobi coordinates was used to 
derive~(\ref{eq:Fourier}) the representation~(\ref{eq:Fourier}) in its general
form can be seen to be valid for any choice of  Jacobi coordinates.
The particular choice~(\ref{eq:Jacobi}) is a preferred choice since it
leads to the specific form of the argument 
$-P^2_1-P^2_2=-\tfrac23(p_1^2+p_2^2+p_1p_2)$.
Since this expression is invariant under the transformations:
$p_1\leftrightarrow p_2$, $p_2\to -p_2-p_1$ and $p_1\to -p_1-p_2$,  
the r.h.s. in Eq.~(\ref{eq:Fourier}) is invariant under permutations 
of all $x_i$ as it should be.

In the next step we have to specify the function 
$\bar\Phi_N(-P^2_1-P^2_2)\equiv\bar\Phi_N(-P^2)$, which 
characterizes the finite size of the baryons.  
We will choose a simple Gaussian form for the function $\bar\Phi_N$
\be
\bar\Phi_N(-P^2) = \exp(P^{\,2}/\Lambda_N^2) \,,
\label{eq:Gauss}
\en  
where $\Lambda_N$ is a size parameter parametrizing the distribution 
of quarks inside a nucleon. Note that we have used another
definition of the $\Lambda_N$ in our previous papers:
 $\Lambda_N = \Lambda_N^{\rm old}/(3\sqrt{2}).$   

Since $P^{\,2}$ turns into $-\,P^{\,2}_E$ in Euclidean space 
the form~(\ref{eq:Gauss}) has the appropriate fall--off behavior in 
the Euclidean region.
We emphasize that any choice for $\Phi_N$ is appropriate
as long as it falls off sufficiently fast in the ultraviolet region of
Euclidean space to render the corresponding Feynman diagrams ultraviolet 
finite. The choice of a Gaussian form for $\Phi_H$ has obvious
calculational advantages.

The coupling constants $g_N$ are determined by 
the compositeness condition suggested by Weinberg~\cite{Weinberg:1962hj}
and Salam~\cite{Salam:1962ap} (for a review, see~\cite{Hayashi:1967hk})
and extensively used by us in previous papers on the covariant quark model  
(for details, see~\cite{Efimov:1993ei}). The compositeness condition 
postulates that the renormalization constant of the bound-state wave function 
is set equal to zero. In the case of a baryon this implies that 
\be 
Z_N = 1 - \Sigma^\prime_N(m_N)  = 0 \,,
\label{eq:Z=0}
\en 
where $\Sigma^\prime_N$ is the on-shell derivative of the nucleon mass 
function $\Sigma_N$, i.e. 
$\Sigma^\prime_N=\partial\,\Sigma_N/\partial\!\!\not\!p$, at 
$p^{2}=m_{N}^{2}$
and where $m_N$ is the nucleon mass. The compositeness condition is the
central equation of our covariant quark model. It can be viewed as the
field theoretic equivalent of the wave function normalization condition
for quantum mechanical wave functions.  
The physical meaning, the implications and corollaries of the compositeness 
condition have been discussed in some detail in our previous papers 
(see e.g.~\cite{Branz:2009cd}). 

\subsection{Infrared confinement}
\label{Sec:IR} 

In~\cite{Branz:2009cd} we have shown how the confinement of 
quarks can be effectively incorporated in the covariant quark model.
In a first step, we introduced an additional scale integration
in the space of Schwinger's $\alpha$--parameters with an integration range
from zero to infinity. In a second step the scale integration was
cut off at the upper limit which corresponds to the introduction of an 
infrared (IR) cutoff. In this manner all possible thresholds 
present in the initial quark diagram were removed. The cutoff parameter was 
taken to be the same for all physical processes. Other model parameters such 
as the constituent quark masses and size parameters were determined from a fit
to experimental data.

Let us describe the basic features of how IR confinement is implemented in
our model. All physical matrix elements are described by Feynman diagrams 
written in terms of a convolution of free quark propagators and the vertex 
functions. Let $n$ and  $m$ be the number of the propagators and 
vertices, respectively. For the current-induced baryon transitions or the
derivative of the mass function one has
four quark propagators and two vertex functions, i.e. one has $n=4$ and 
$m=2$. In Minkowski space the two-loop diagram will be represented as
\bea
\Pi(p_1,...,p_m) &=& 
\int\!\! [d^4k]^2  
%\int\!\! [d^4k]^\ell  
\prod\limits_{i_1=1}^{m} \,
\Phi_{i_1+n} \left( -K^2_{i_1+n}\right)
\prod\limits_{i_3=1}^n\, S_{i_3}(\tilde k_{i_3}+v_{i_3}),
\nn
&&\nn
K^2_{i_1+n} &=& \sum_{i_2}(\tilde k^{(i_2)}_{i_1+n}+v^{(i_2)}_{i_1+n})^2\,,
\label{eq:diag}
\ena
where the vectors $\tilde k_i$  are  linear combinations of the loop momenta 
$k_i$. The $v_i$ are  linear combinations of the external momenta $p_i$. 
The strings of Dirac matrices appearing in the calculation need not concern 
us here since they do not depend on the momenta. 
The external momenta $p_i$ are all chosen to be ingoing such that one has 
$\sum\limits_{i=1}^m p_i=0$. 
 
Using the Schwinger representation the local quark propagator is written as 
\be
S(k) = \frac{1}{m -\not\! k}=(m +\not\! k)
\int\limits_0^\infty\! 
d\beta\,e^{-\beta\,(m^2-k^2)}\,.
\en
As mentioned before one takes the Gaussian form or the vertex functions, i.e.
\be
\label{eq:vert} 
\bar{\Phi}_{i+n} \left( -K^2\right)\,
 =
\exp\left[\beta_{i+n}\,K^2\right] \qquad i=1,...,m\, ,
\en
where, as in~(\ref{eq:Gauss}), the parameters $\beta_{i+n}$ are related to 
the respective size 
parameters of the baryons $\Lambda_{i}$ via $\beta_{i+n}=1/\Lambda^2_{i}$. 
The integrand in Eq.~(\ref{eq:diag})
has a Gaussian form with the exponential factor $(kak+2kr+R)$ where,
in case of the baryonic two-loop calculation, $a$ is a
$2\times 2$ matrix depending on the parameters $\beta_i$,
$r$ is a $2$-component vector composed from the external momenta,
$k$ is a  $2$-component vector of the loop momenta of the form 
$k = (k_1,k_2)$ and  
$R$ is a quadratic form of the external momenta.
Tensor loop integrals are calculated with the help of the differential
representation 
\be
k_i^\mu e^{2kr} = \frac{1}{2}\frac{\partial}{\partial r_{i\,\mu}}e^{2kr}\,.
\en 
After doing the loop integration the differential operators 
$\partial/\partial r_{i\,\mu}$ will give cause to outer momenta tensors 
which, in the present case, are $p$ and $p'$. 
We have written a FORM~\cite{Vermaseren:2000nd} program that achieves 
the necessary commutations of the differential operators in a very efficient 
way.

After doing the loop integrations one obtains 
($n$ denotes the number of propagators)
\be
\Pi =  \int\limits_0^\infty d^n \beta \, F(\beta_1,\ldots,\beta_n) \,,
\en
where $F$ stands for the whole structure of a given diagram. 
The set of Schwinger parameters $\beta_i$ can be turned into a simplex by 
introducing an additional $t$--integration via the identity 
\be 
1 = \int\limits_0^\infty dt \, \delta(t - \sum\limits_{i=1}^n \beta_i)
\en 
leading to 
\be
\hspace*{-0.2cm}
\Pi   = \int\limits_0^\infty\! dt t^{n-1}\!\! \int\limits_0^1\! d^n \alpha \, 
\delta\Big(1 - \sum\limits_{i=1}^n \alpha_i \Big) \, 
F(t\alpha_1,\ldots,t\alpha_n). 
\label{eq:loop_2} 
\en
There are altogether $n$ numerical integrations: $(n-1)$ $\alpha$--parameter
integrations and the integration over the scale parameter $t$. 
The very large $t$--region corresponds to the region where the singularities
of the diagram with its local quark propagators start appearing. 
However, as described in \cite{Branz:2009cd}, if one introduces 
an IR cutoff on the upper limit of the $t$--integration, all 
singularities vanish because the integral is now analytic for any value
of the set of kinematic variables.
We cut off the upper integration at $1/\lambda^2$ and obtain
\be
\label{eq:IR-cutoff}
\hspace*{-0.2cm}
  \Pi^c = \!\!  
\int\limits_0^{1/\lambda^2}\!\! dt t^{n-1}\!\! \int\limits_0^1\! d^n \alpha \, 
\delta\Big(1 - \sum\limits_{i=1}^n \alpha_i \Big) \, 
F(t\alpha_1,\ldots,t\alpha_n).
\en  
By introducing the IR cutoff one has removed all potential thresholds 
in the quark loop diagram, i.e. the quarks are never on-shell and are thus
effectively confined. We mention that an explicit demonstration of the 
absence of a two- quark threshold in the case of a scalar one--loop two--point 
function has been given in Ref.~\cite{Branz:2009cd}.
We take the infrared cutoff parameter $\lambda$ to be the 
same in all physical processes. 
The numerical evaluations have been done by a numerical program 
written in the FORTRAN code.
%-----------------------------------------------------------------------------
\section{Electromagnetic interactions}
%-----------------------------------------------------------------------------
We use the standard free fermion Lagrangian for the 
baryon and quark fields: 
\be
{\cal L}_{\rm free}(x) =  
\bar B(x) (i \not\!\partial - m_B) B(x) + 
\sum\limits_q \, \bar q(x) (i \not\!\partial - m_q) q(x) \, , 
\label{eq:Lagr_free}
\en 
where $m_q$ is the constituent quark mass. 
The interaction with the electromagnetic field has to be introduced both at 
the baryon and the quark level. In a first step we gauge the free 
Lagrangians Eq.~(\ref{eq:Lagr_free}) of the quark 
and baryon fields in the standard manner by using minimal
substitution:
\be
\partial^\mu B \to (\partial^\mu - ie_B A^\mu) B\,, 
\qquad
\partial^\mu q_i \to (\partial^\mu - ie_{q_i} A^{\mu}) q_i\,, 
 \label{eq:em_min} 
\en 
where $e_B$ is the electric charge of the baryon $B$ and 
$e_{q_i}$ is the electric charge of the quark with flavor 
$q_i$. 
The interaction of the baryon and quark fields with the e.m. field is thus 
specified by minimal substitution. 
The interaction Lagrangian reads
\be 
{\cal L}^{\rm em-min}_{\rm int}(x) = 
e_B \bar B(x) \!\not\!\! A \, B(x)  +  
\sum\limits_q \, e_q \, \bar q(x) \!\not\!\! A  \, q(x) \,. 
\label{eq:Lagr_em_min}
\en 
As will become apparent further on, the electromagnetic field does not directly
couple to the baryon fields as a result of the compositeness condition.

Next one gauges the nonlocal Lagrangian Eq.~(\ref{eq:Lagr_str}).
The gauging proceeds in a way suggested in 
Refs.~\cite{Mandelstam:1962mi,Terning:1991yt}
and used before by us (see, for instance,
Refs.~~\cite{Ivanov:1996pz,Faessler:2006ft}).
In order to guarantee local invariance of the strong interaction 
Lagrangian one multiplies 
each quark field $q(x_i)$ in ${\cal L}_{\rm int}^{\rm str}$ with a 
gauge field exponential.
One then has
\be
q_i(x_i)\to e^{-ie_{q_1} I(x_i,x,P)} \, q_i(x_i)\,,
\label{eq:gauging}
\en 
where
\be
I(x_i,x,P) = \int\limits_x^{x_i} dz_\mu A^\mu(z). 
\label{eq:path}
\en
The path $P$ connects the end-points of the path integral.

It is readily seen that the full Lagrangian is invariant 
under the transformations
\bea 
\begin{aligned}
   &  q_i(x) \, \to \, e^{ie_{q_i} f(x)} q_i(x)\,,   \qquad
 \bar q_i(x) \, \to \, \bar q_i(x) \, e^{-ie_{q_i} f(x)}\,, 
\nn
   & B(x) \, \to \, e^{ie_B f(x)} \, B(x)\,, \qquad 
\bar B(x) \, \to \, \bar B(x) \, e^{-ie_B f(x)}\,, 
\nn 
   & A^\mu(x) \, \to \, A^\mu(x)+\partial^\mu f(x)\,, 
\label{eq:gauge_groop}
\end{aligned}
\ena
where $e_B = \sum\limits_{i=1}^3 e_{q_i}$. 

One then expands the gauge exponential
up to the requisite power of $e_{q}A_\mu$ needed in the perturbative series. 
This will give rise to a second term in the nonlocal electromagnetic 
interaction Lagrangian ${\cal L}^{\rm em-nonloc}_{\rm int}$. 
Superficially it appears that the results will depend on the path $P$
taken to connect the end-points in the path integral in 
Eq.~(\ref{eq:path}).  
However, one needs to know only the derivatives of the path integral
expressions when calculating the perturbative series.
Therefore, we use the 
formalism suggested in~\cite{Mandelstam:1962mi,Terning:1991yt} 
which is based on the path-independent definition of the derivative of 
$I(x,y,P)$: 
\be
\lim\limits_{dx^\mu \to 0} dx^\mu 
\frac{\partial}{\partial x^\mu} I(x,y,P) \, = \, 
\lim\limits_{dx^\mu \to 0} [ I(x + dx,y,P^\prime) - I(x,y,P) ]
\label{eq:path1}
\en 
where the path $P^\prime$ is obtained from $P$ by shifting the end-point $x$
by $dx$.
The definition (\ref{eq:path1}) leads to the key rule
\be
\frac{\partial}{\partial x^\mu} I(x,y,P) = A_\mu(x)
\label{path2}
\en 
which in turn states that the derivative of the path integral $I(x,y,P)$ does 
not depend on the path $P$ originally used in the definition. 

As a result of this rule the Lagrangian describing the nonlocal interaction
of the baryon, quark and electromagnetic fields to the first order in the 
electromagnetic charge reads
\be
{\cal L}^{\,\rm em-nonloc}_{\rm int}(x) = 
  g_B \,\bar B(x)\cdot\!\!\int\!\! dy A_\alpha(y) J_{B-\rm em}^{\alpha}(x,y) 
+ g_B \int\!\! dy A_\alpha(y) \bar J_{B-\rm em}^{\alpha}(x,y)\cdot B(x)\,,  
\label{eq:Lagr_em-nonloc}
\en
where the nonlocal electromagnetic currents are given by
\bea
J_{B-\rm em}^{\alpha}\,(x,y) &=& \prod\limits_{i=1}^3\int\!\! dx_i  
\, J^{(B)}_{3q}(x_1,x_2,x_3)\,E^\alpha_B(x;x_1,x_2,x_3;y)\,,
\nn
\bar J_{B-\rm em}^{\alpha}(x,y) &=& \prod\limits_{i=1}^3\int\!\! dx_i  
\, \bar J^{(B)}_{3q}(x_1,x_2,x_3)\,E^{\alpha\,\dagger}_B(x;x_1,x_2,x_3;y)\,,
\nn
\nn
E^\alpha_B(x;x_1,x_2,x_3;y) &=& 
\prod\limits_{i=1}^3\int\!\! \frac{dp_i}{(2\pi)^4}\!\! 
\int\!\! \frac{dr}{(2\pi)^4}\,
e^{-i\sum\limits_{i=1}^3p_i(x-x_i)-ir(x-y)}\,
\widetilde E^\alpha_B(p_1,p_2,p_3;r)\,,
\nn
\widetilde E^\alpha_B(p_1,p_2,p_3;r) &=& 
\sum\limits_{i=1}^3e_{q_i}\int\limits_0^1 d\tau
\Big\{- w_{i1}(w_{i1}r^\alpha + 2 q_1^\alpha) \bar\Phi'_B(-z_1)
      - w_{i2}(w_{i2}r^\alpha + 2 q_2^\alpha) \bar\Phi'_B(-z_2)\Big\}\,,
\label{eq:cur-em-nonloc}
\ena
Further
\bea
q_1 &=& \sum\limits_{i=1}^3 w_{i1}p_i\,, \qquad
q_2 = \sum\limits_{i=1}^3 w_{i2}p_i \,,
\nn
z_1 &=&\tau(w_{i1}r+q_1)^2-(1-\tau)q_1^2- (w_{i2}r+q_2)^2\,,\qquad
z_2= q_1^2-\tau(w_{i2}r+q_2)^2-(1-\tau)q_2^2\,,
\nn
\nn
w_{i1} &=& \left(\begin{array}{c}
 \tfrac{1}{\sqrt{2}} w_3   \\
 \tfrac{1}{\sqrt{2}} w_3   \\
-\tfrac{1}{\sqrt{2}} (w_1+w_2) \end{array}\right)\,,\qquad
w_{i2} = \left(\begin{array}{c}
-\tfrac{1}{\sqrt{6}} (2w_2 + w_3)  \\
 \tfrac{1}{\sqrt{6}} (2w_1+w_3)   \\
 \tfrac{1}{\sqrt{6}} (w_1-w_2) \end{array}\right)\,.
\ena

\section{Nucleon mass function and electromagnetic form factors}

We start with the calculation of the proton mass function (also called 
self-energy function) needed for the implementation of the compositeness 
condition. The relevant term in the expansion of the $S$--matrix reads
\bea
\label{massfun}
S_2 &=& i^2g_N^2 \int\!\! dx \!\!  \int \!\! dy \,
\bar p(x) \la 0|T\{J_p(x)\bar J_p(y)\}|0\ra p(y)\,
\nn
&\doteq& i  \int\!\!  dx\!\!  \int\!\!  dy \,\bar p(x)\Sigma_p(x-y) p(y) \,. 
\label{eq:self1}
\ena 
The corresponding two--loop Feynman quark diagram is shown in Fig.~1. 
In Eq.~(\ref{massfun}) we have introduced the standard notation for the proton
mass function 
\be
\Sigma_p(x-y) = ig_N^2\, \la 0|T\{J_p(x)\bar J_p(y)\}|0 \ra\,.
\label{eq:self2}    
\en
The Fourier-transform of the mass function $\Sigma_p(x-y)$ is given
by
\be
\Sigma_p(x-y) = \int\!\! \frac{d^4p}{(2\pi)^4}\,e^{-ip(x-y)}\,
\Sigma_p(p)\,, \qquad 
\Sigma_p(p) = \int\!\!  d^4x\,e^{ipx}\Sigma_p(x)\,.
\label{eq:self-Fourier}    
\en 

We use the same notation $\Sigma_{p}$ for the mass function of the
proton in coordinate space and in momentum space. Which of the two 
representations
are being used can be read off from the arguments, {\it cif.} $\Sigma_p(x-y)$
and $\Sigma_p(p)$.  The matrix element of $S_2$ in Eq.~(\ref{eq:self1}) 
between the initial and final proton states (with momenta
$p$ and $p'$, respectively) is expressed by
\bea
\la p'|S_2|p \ra \,=\, i\,(2\pi)^4\,\delta^{(4)}(p-p')\,
\bar u_p(p)\,\Sigma_p(p)\,u_p(p)\,.
\label{eq:self3}
\ena

It is straightforward but nevertheless cumbersome to calculate the proton 
mass function $\Sigma_p(x-y)$. One uses the explicit 
expression for the interpolating  three--quark current given by 
Eq.~(\ref{eq:current}) and the time--ordering of quark fields:
\be
\la 0|T\{\,q^a_f(x)\bar q^{\,a'}_{f'}(y)\,\}|0 \ra = 
\delta_{aa'}\delta_{ff'}\,S_f(x-y)
=\delta_{aa'}\delta_{ff'}\int\!\!\frac{d^4k}{(2\pi)^4i}\,e^{-ik(x-y)}\, 
S_f(k)\,,
\label{eq:Green-quark}
\en
where $S_f(x-y)$ and $S_f(k)$ are the free quark propagators in 
coordinate and momentum space with
\be 
S_f(k)  = \frac{1}{m_{f}-\not\! k}\,
\label{eq:Green-quark2}
\en
and where $a,a'$ and $f,f'$ are color and flavor indices, respectively.  

In momentum space the proton mass function is given by 
\be
\Sigma_p(p) = 12 g_N^2 \,\int\!\!\frac{d^4k_1}{(2\pi)^4i}\!\!\int\!\!
\frac{d^4k_2}{(2\pi)^4i}\bar\Phi_N^2(-z_0)
\Gamma^A\gamma^5 S_d(k_1+w_1 p)\gamma^5\Gamma^B
\Tr\left[S_u(k_2-w_2 p)\Gamma_A S_u(k_2-k_1+w_3 p)\Gamma_B\right]\,,
\label{eq:mass}
\en
where 
\be
z_0 =\tfrac12(k_1-k_2)^2+\tfrac16(k_1+k_2)^2\,.
\label{eq:short2}
\en
In order to economize on the notation we introduce a short-hand notation for 
the two loop momentum integrations in~(\ref{eq:mass}) and
in the following formulas. We write
\be
\la\la \ldots \ra\ra = 
\int\!\!\frac{d^4k_1}{(2\pi)^4i}\!\!\int\!\!\frac{d^4k_2}{(2\pi)^4i}
\,(\ldots)\,\,.
\label{eq:short1}
\en

Note that the integral~Eq.~(\ref{eq:mass}) is invariant under a shift
of the loop momenta $k_i\to k_i + a p$ where $a$ is an arbitrary number and
$p$ is the outer momentum. Using this invariance one can obtain various 
equivalent representations for the mass
function. In Eq.~(\ref{eq:mass}) we have chosen $a$ such that the external 
momentum does not appear in the argument of the vertex function.
One has $w_2=w_3=\tfrac12(1-w_1)$ where $w_1=m_d/(m_d+2m_u)$.
We mention that it is convenient to keep $m_u \neq m_d$ in the analytical 
calculation in order to distinguish the proton from the neutron case. 
In the end, when we do the numerical calculation, we set $m_u=m_d$. 

According to the compositeness condition Eq.~(\ref{eq:Z=0}) one needs to
calculate the derivative of the proton mass function. Since the 
proton is on mass--shell, i.e.
$\bar u(p)\!\!\!\not\!\!p \,u(p) =m_N \bar u(p) u(p)$ and hence 
$p^\mu \bar u(p) u(p) = m_N \bar u(p) \gamma^\mu u(p)$, the compositeness
condition 
\be
\Sigma_p'(p)=
\frac{\partial\Sigma_p(p)}{\partial\!\!\not\!p}\,=\,1
\quad \text{with}\quad \not\!p=m_N
\en
can be written as
\be
\frac{\partial\Sigma_p(p)}{\partial p_\mu}\,=\,\gamma^\mu  
\quad \text{with}\quad  \not\!p=m_N \quad \text{and}\quad
 p^\mu=m_N\gamma^\mu\,.
\label{eq:mass-deriv1}
\en 
Here and in the following it is understood that the relations  
between Green functions are valid when sandwiched between 
spinors.  
The latter form~(\ref{eq:mass-deriv1}) is more suitable for our
calculation because of its relation to the electromagnetic proton vertex 
function
at zero momentum transfer. Using
\be
\frac{\partial}{\partial p_\mu}S_{f}(k+w p)=w\,S_{f}(k+w p)\gamma^\mu 
S_{f}(k+w p)
\en
one obtains
\bea
\frac{\partial\Sigma_p(p)}{\partial p_\mu} 
&=& 12 g_N^2
{\Bla\Bla}\bar\Phi_N^2(-z_0)
\Big\{
w_1 \Gamma^A\gamma^5 S_d(k_1+w_1 p)\gamma^\mu S_d(k_1+w_1 p)\gamma^5\Gamma^B
\Tr\left[S_u(k_2-w_2 p)\Gamma_A S_u(k_2-k_1+w_3 p)\Gamma_B\right]
\nn
&&
-\,w_2 \Gamma^A\gamma^5 S_d(k_1+w_1 p)\gamma^5\Gamma^B
\Tr\left[S_u(k_2-w_2 p)\gamma^\mu S_u(k_2-w_2 p)\Gamma_A 
S_u(k_2-k_1+w_3 p)\Gamma_B\right]
\nn
&&
+\,w_3 \Gamma^A\gamma^5 S_d(k_1+w_1 p)\gamma^5\Gamma^B
\Tr\left[S_u(k_2-w_2 p)\Gamma_A S_u(k_2-k_1+w_3 p)\gamma^\mu
S_u(k_2-k_1+w_3 p)\Gamma_B\right]
\Big\}\Bra\Bra\,.
\label{eq:mass-deriv2}
\ena

We now return to the calculation of the electromagnetic vertex of the proton. 
There are two terms in the relevant expansion of the $S$-matrix. 
These are derived i) from the Lagrangian Eq.~(\ref{eq:Lagr_em_min}) describing
the local interaction of the photon with the quarks and ii) from the 
Lagrangian 
Eq.~(\ref{eq:Lagr_em-nonloc}) describing the nonlocal interaction 
nucleon+quarks+photon. One has
\bea
S_3 &=& i^3g_N^2 \int\!\! dx \!\!  \int \!\! dy\!\!  \int \!\! dz \,
A_\mu(z)\, \bar p(x) \la 0|T\{J_p(x)\,
\left(  e_u\,\bar u(z)\gamma^\mu u(z) + e_d\,\bar d(z)\gamma^\mu d(z) \right)
\bar J_p(y)\}|0\ra p(y)
\nn
&+&  i^2g_N^2 \int\!\!  dx\!\!  \int\!\!  dy\!\!  \int \!\! dz \,
A_\mu(z)\,\bar p(x) \la 0|T\{ J_p(x)\,\bar J_{p-\rm em}^\mu(y,z)
                       + J_{p-\rm em}^\mu(x,z) \bar J_p(y) \}|0\ra  p(y)\,,
\label{eq:S3}
\ena
where the currents $J_p (\bar J_p)$ and 
$J_{p-\rm em}^\mu (\bar J_{p-\rm em}^\mu)$
are defined by Eqs.~(\ref{eq:current}) and~(\ref{eq:cur-em-nonloc}),
respectively. It is important to keep track of the signs of the various
charges in the calculation. Our choice is to take the electric charges
of charged particles in units of the proton charge, e.g.
$e_p=+1$, $e_u=+2/3$, $e_d=-1/3$, etc.

The matrix element $S_3$ in Eq.~(\ref{eq:S3}) taken 
between the initial and final proton states with momenta ($p$) and ($p'$)
and the photon state with momentum ($q=p-p'$) reads
\be
\la p';q,\mu|S_3|p\ra \,=\,
\bar u_p(p')\,T_3^\mu(p;p',q)\,u_p(p)\,.
\label{eq:T3}
\en
The matrix element $T^\mu_3(p;p'q)$ is obtained from Eq.~(\ref{eq:S3})
by the substitutions $A_\mu(z)\to e^{iqz}$, $\bar p(x)\to e^{ip'x}$
and $p(y)\to e^{-ipy}$.
A straightforward calculation gives 
\be
T_3^\mu(p;p',q)= i(2\pi)^4\delta^{(4)}(p-p'-q)\Lambda_p^\mu(p,p')\,
\en
where the electromagnetic vertex function $\Lambda_p^\mu(p,p')$ of the proton
consists of four pieces represented by the four two-loop quark diagrams 
in Fig.2:
\begin{itemize}
\item the vertex diagram with the e.m. current attached to the d-quark (Fig.2a) \ 
--- $\Lambda^\mu_{p\,\rm d}$,
\item the vertex diagram with the e.m. current attached to the u-quark (Fig.2b) \ 
--- $\Lambda^\mu_{p\,\rm u}$,
\item two bubble diagrams with the e.m. current attached to the initial proton vertex
(Fig.2c) \ 
--- $\Lambda^\mu_{p\, (a)}$, \\
and with the e.m. current attached to the final proton vertex
(Fig.2d) \ 
--- $\Lambda^\mu_{p\, (b)}$.
\end{itemize}

The four different contributions can be calculated to be 

\bea
\Lambda^\mu_{p\,\rm d}(p,p')
&=& -\,4\, g_N^2
\Bla\Bla\bar\Phi_N(-z_0)
  \bar\Phi_N
\Big(-\tfrac12(k_1-k_2+w_3 q)^2-\tfrac16(k_1+k_2+(2w_2+w_3) q)^2\Big)
\nn
&\times&
\Gamma^A\gamma^5 S_d(k_1+w_1 p')\gamma^\mu S_d(k_1+w_1 p'+q)\gamma^5\Gamma^B
\Tr\left[S_u(k_2-w_2 p')\Gamma_A S_u(k_2-k_1+w_3 p')\Gamma_B\right]
\Bra\Bra\,,
\nn\nn
\Lambda^\mu_{p\,\rm u}(p,p')
&=& 16\, g_N^2
\Bla\Bla\bar\Phi_N(-z_0)
  \bar\Phi_N\Big(-\tfrac12(k_1-k_2-(w_1+w_2) q)^2
                 -\tfrac16(k_1+k_2-(w_1-w_2) q)^2\Big)
\nn
&\times&
\Gamma^A\gamma^5 S_d(k_1+w_1 p')\gamma^5\Gamma^B
\Tr\left[S_u(k_2-w_2 p')\Gamma_A 
S_u(k_2-k_1+w_3 p')\gamma^\mu S_u(k_2-k_1+w_3 p'+q)\Gamma_B\right]\Bra\Bra\,,
\nn\nn
\Lambda^\mu_{p\,(a)}(p,p')
&=& 12\, g_N^2
\Bla\Bla\bar\Phi_N(-z_0)
\tilde E_p^\mu(k_1+w_1p',-k_2+w_2p',k_2-k_1+w_3p';q)
\nn
&\times&
\Gamma^A\gamma^5 S_d(k_1+w_1 p')\gamma^5\Gamma^B
\Tr\left[S_u(k_2-w_2 p')\Gamma_A S_u(k_2-k_1+w_3 p')\Gamma_B\right]\Bra\Bra\,,
\nn\nn
\Lambda^\mu_{p\, (b)}(p,p')
&=& 12\, g_N^2
\Bla\Bla\bar\Phi_N(-z_0)
\tilde E_p^\mu(k_1+w_1p,-k_2+w_2p,k_2-k_1+w_3p;-q)
\nn
&\times&
\Gamma^A\gamma^5 S_d(k_1+w_1 p)\gamma^5\Gamma^B
\Tr\left[S_u(k_2-w_2 p)\Gamma_A S_u(k_2-k_1+w_3 p)\Gamma_B\right]\Bra\Bra\,.
\label{eq:em-vertex}
\ena
The two bubble diagrams Figs.~2c-2d can be seen to be related by
\be
\Lambda^\mu_{p\,(b)}(p,p')=\Lambda^\mu_{p\,(a)}(p',p)\,.
\label{eq:bubble0}
\en

Gauge invariance requires the validity of the $q^{\mu}=0$ Ward identity  
\be
\frac{\partial\Sigma_p(p)}{\partial p_\mu} = \Lambda_p^\mu(p,p)\,,
\label{eq:WI}
\en
where $\Lambda_p^\mu(p,p)$ is given by the sum of the four contributions
Eq.~(\ref{eq:em-vertex}).
The l.h.s. of~(\ref{eq:WI}) has been written down in 
Eq.~(\ref{eq:mass-deriv2}). The contribution
of the bubble diagrams to the r.h.s. of~(\ref{eq:WI}) is calculated by using
the explicit representation of $\tilde E^\mu_p$ in 
Eq.~(\ref{eq:cur-em-nonloc}). For the proton the quark charges are given by 
$e_1=e_d$ and $e_2=e_3=e_u$. One finds
\bea
\Lambda^\mu_{p\,(a)}(p,p)+\Lambda^\mu_{p\,(b)}(p,p)
&=&\, 8\,(1+3w_1)\,g_N^2
\la\la k_1^\mu  \bar\Phi'_N(-z_0)\bar\Phi_N(-z_0)
\nn
&\times&
\Gamma^A\gamma^5 S_d(k_1+w_1 p)\gamma^5\Gamma^B
\Tr\left[S_u(k_2-w_2 p)\Gamma_A S_u(k_2-k_1+w_3 p)\Gamma_B\right]\ra\ra\,.
\label{eq:bubble}
\ena

Superficially such terms are not present in Eq.~(\ref{eq:mass-deriv2})
since~(\ref{eq:mass-deriv2}) contains four quark propagators as a result
of having differentiated the vertex function as compared to the three
propagators in~(\ref{eq:bubble}).
However, the bubble contributions~(\ref{eq:bubble}) may be rewritten in terms
of the vertex diagram contributions at $q=0$. This is achieved by using an 
integration-by-parts (IBP) identity where on differentiates the integrand 
w.r.t. the loop momentum $k_{1}$. One has 
\be
\Bla\Bla \, \frac{\partial}{\partial k_1^\mu}
\Big\{\bar\Phi^2_N(-z_0)
\Gamma^A\gamma^5 S_d(k_1+w_1 p)\gamma^5\Gamma^B
\Tr\left[S_u(k_2-w_2 p)\Gamma_A S_u(k_2-k_1+w_3 p)\Gamma_B\right]
\Big\} \, \Bra\Bra\, \equiv\, 0\,.
\label{eq:IP}
\en

Upon differentiation and use of the symmetry of the integrand
under $k_2\to -k_2 +k_1$ one obtains

\bea
&&
2\,\Bla\Bla k_1^\mu\,\bar\Phi'_N(-z_0)\,\bar\Phi_N(-z_0)
\Gamma^A\gamma^5 S_d(k_1+w_1 p)\gamma^5\Gamma^B
\Tr\left[S_u(k_2-w_2 p)\Gamma_A S_u(k_2-k_1+w_3 p)\Gamma_B\right]\Bra\Bra 
\nn
&=&
\Bla\Bla\bar\Phi^2_N(-z_0)
\Gamma^A\gamma^5 S_d(k_1+w_1 p)\gamma^\mu S_d(k_1+w_1 p)\gamma^5\Gamma^B
\Tr\left[S_u(k_2-w_2 p)\Gamma_A S_u(k_2-k_1+w_3 p)\Gamma_B\right]\Bra\Bra 
\nn
&-&
\Bla\Bla\bar\Phi^2_N(-z_0)
\Gamma^A\gamma^5 S_d(k_1+w_1 p)\gamma^5\Gamma^B
\Tr\left[S_u(k_2-w_2 p)\Gamma_A 
S_u(k_2-k_1+w_3 p)\gamma^\mu S_u(k_2-k_1+w_3 p)\Gamma_B\right]\Bra\Bra\,.
\label{eq:identity}
\ena
Using the IBP-identity and summing up all contributions on the r.h.s.
of Eq.~(\ref{eq:WI}) one finds agreement with the l.h.s. of Eq.~(\ref{eq:WI})
as given by Eq.~(\ref{eq:mass-deriv2}).
One has thus proven the validity of the Ward identity (\ref{eq:WI})
(recall that $w_2=w_3=1/2 (1-w_1)$).

A further technical remark is in order concerning the above check of
the Ward identity Eq.~(\ref{eq:WI}). The proof made use
of an IBP identity assuming the vanishing of the pertinent surface term.
However, in the confinement ansatz with the accompanying IR 
cutoff the requisite surface terms no longer vanish. 
As it turns out one can avoid the use of IBP identities in the proof of the 
Ward identity by astutely shifting the loop momentum $k_{1}$ in the mass
function. The appropriate shift is 
$ k_1\to k_1+(w_2+w_3-\tfrac43) p $ and $ k_2\to k_2+w_2p $. 
One then obtains
\be
\Sigma_p(p) = 
12 g_N^2 \,\int\!\!\frac{d^4k_1}{(2\pi)^4i}\!\!\int\!\!\frac{d^4k_2}{(2\pi)^4i}
\bar\Phi_N^2(-z_1)
\Gamma^A\gamma^5 S_d(k_1 - \tfrac13 p)\gamma^5\Gamma^B
\Tr\left[S_u(k_2)\Gamma_A S_u(k_2-k_1+ \tfrac43 p)\Gamma_B\right]\,,
\label{eq:mass1}
\en
where 
\be
z_1 =\tfrac12 (k_1 - k_2 +(w_3 - \tfrac43) p)^2
    + \tfrac16( k_1 + k_2 + (2 w_2 + w_3 -\tfrac43) p)^2\,.
\label{eq:short3}
\en
After the shift of the loop momentum the argument of the vertex function 
$\bar\Phi_N(-z_1)$ now depends on the external momentum $p$.
When differentiating w.r.t. the external momentum $p$
a new term will appear caused by the derivative of the vertex function 
$\bar\Phi_N$ in addition to the terms originating from the derivatives of the 
quark propagators. After shifting back the loop momenta $z_1\to z_0$
and some algebraic juggling one finds that the derivative of the mass function
coincides analytically with the expression for the electromagnetic vertex
function $\Lambda_p^\mu(p,p)$ given by the sum of the contributions
of the triangle diagrams~(\ref{eq:em-vertex}) and the bubble diagrams
Eq.~(\ref{eq:bubble}). We emphasize that in this derivation we did not have 
to make use of an IBP identity to prove the Ward identity.

Hereafter we will use the compositeness condition $Z_N=0$ 
in the form 
\be
  \Lambda^\mu_{p\,d}(p,p)   + \Lambda^\mu_{p\,u}(p,p)
+ \Lambda^\mu_{p\,(a)}(p,p)  + \Lambda^\mu_{p\,(b)}(p,p) = \gamma^\mu
\qquad\text{with}\quad (\not\! p=m_N) 
\quad \text{and}\quad p^\mu=m_N\gamma^\mu
\label{eq:gN}
\en
in order to determine the coupling constant $g_N$. 
This allows one to provide the correct normalization
of the charged proton form factor within the confinement scenario.

Another useful check is to reproduce the generalized Ward-Takahashi identity 
\be
q_\mu\Lambda_p^\mu(p,p')=\Sigma_p(p)-\Sigma_p(p')\,.
\label{eq:GWI}
\en
We shall not elaborate on this proof which is straightforward by using 
suitable shifts of the loop variables.

Let us briefly describe another check on the gauge invariance of our 
calculation. Without gauge invariance there are three independent Lorentz 
structures in the electromagnetic proton vertex which can be chosen to be
\be
\Lambda_p^\mu(p,p')=\gamma^\mu\,F^p_1(q^2) 
- \frac{i\sigma^{\mu q}}{2m_N}\,F^p_2(q^2)
+q^\mu\,F^p_{NG}(q^2)\,,
\label{eq:em-ff}
\en
where $\sigma^{\mu q} = 
\tfrac{i}{2}(\gamma^\mu\gamma^\nu-\gamma^\nu\gamma^\mu)q_\nu.$
The form factor $F^p_{NG}(q^2)$ characterizes the non--gauge invariant piece
and must therefore vanish for any $q^2$ in a calculation which respects
gauge invariance. For the four contributions of Fig.~2a-2d we found that 
\be
F^p_{NG\, \rm d}(q^2) \equiv 0\,, \qquad  
F^p_{NG\, \rm u}(q^2) \equiv 0\,, \qquad 
F^p_{NG\, (b)}(q^2)   \equiv -\,F^p_{NG\, (a)}(q^2) \qquad \forall q^2.
\label{eq:NG-vanish}
\en  
The gauge variant contributions of the two vertex diagrams are zero while
they vanish for the sum of the two bubble diagrams. 

Before discussing the e.m. properties of the neutron
we would like to comment on a potential conflict between gauge invariance
and our confinement ansatz.
In general the IR cutoff used in Eq.~(\ref{eq:IR-cutoff})
can destroy the gauge invariance as any cutoff can do.
One can, however, show that in some special cases
gauge invariance remains
unimpaired when implementing confinement through an IR cutoff.
For example, in Appendix B
of~\cite{Branz:2009cd} we have shown that the $\rho-\gamma$ 
transition amplitude is gauge invariant off mass-shell even in the presence
of an IR cutoff. The crucial point of the proof was that we were able to show 
that the integrand of the t-integration itself was gauge invariant. 
In the case of the electromagnetic form factor of the proton one finds again
that the integrand of the t-integration is gauge invariant by itself
due to a symmetry property of the integrand in the space of the Schwinger 
$\alpha$-parameters.  
However, if the proof of gauge invariance requires
an integration by parts in the space of momenta which becomes
translated into an integration by parts over the t-parameter
gauge invariance will be spoiled by the surface term due to the
upper integration limit $1/\lambda^2$. In order to keep
gauge invariance one can proceed as follows. First, by using the properties
of the relevant integrals over the loop momenta one needs to
specify a gauge invariant part of the full amplitude.
Then one employs our confinement ansatz for the gauge invariant parts of the 
amplitudes. Such an approach was used to verify the validity of the Ward 
identity when connecting the derivative
of the mass function and the electromagnetic vertex function in the presence
of an IR-cutoff. 

The electromagnetic vertex function of the neutron
is obtained from that of the proton by replacing $m_u\leftrightarrow m_d$,
$e_u\leftrightarrow e_d$ and $m_p\to m_n$. $F_1^N(q^2)$ and 
$F_2^N(q^2)$ are the Dirac and Pauli nucleon form factors 
which are normalized to the electric charge $e_N$ and anomalous 
magnetic moment $k_N$ ($k_N$ is given in units of the nuclear magneton 
$e/2m_p$),respectively, i.e. one has $F_1^N(0)=e_N$ and $F_2^N(0)=k_N$.  
In particular, one can analytically check by using the IBP identity 
that the Dirac form factor of the neutron is equal to zero at $q^2=0$.
  
The nucleon magnetic moments $\mu_N = F_1^N(0)+F_2^N(0)$ 
are known experimentally with high accuracy~\cite{Nakamura:2010zzi}
\be
\mu^{\rm expt}_p = 2.79 \qquad 
\mu^{\rm expt}_n=-1.91 \,.
\label{eq:mag-mom-expt}
\en
We will use these values to fit the value of the nucleon size parameter.
The other model parameters are taken from the fit to mesonic transitions 
done in~\cite{Ivanov:2011aa}:
\be
\def\arraystretch{2}
\begin{array}{cccccc}
     m_u        &      m_s        &      m_c       &     m_b & \lambda  &   
\\\hline
 \ \ 0.235\ \   &  \ \ 0.424\ \   &  \ \ 2.16\ \   &  \ \ 5.09\ \   & 
\ \ 0.181\ \   & \ {\rm GeV} 
\end{array}
\label{eq: fitmas}
\en
We obtain 
\bea
\text{vector current} &\Longrightarrow&
\Lambda_N = 0.36\,  \text{GeV}\, \quad \mu_p = 2.79 \quad  \mu_n = -1.70\,,
\\[2ex]
\text{tensor current} &\Longrightarrow&
\Lambda_N = 0.61\,  \text{GeV}\, \quad \mu_p = 2.79 \quad  \mu_n = -1.69\,.
\label{eq:mag-mom}
\ena

It is convenient to introduce the Sachs electromagnetic form factors 
of nucleons 
\bea
G_E^N(q^2) &=& F_1^N(q^2) + \frac{q^2}{4m^2_N} F_2^N(q^2)\,,
\nn
G_M^N(q^2) &=& F_1^N(q^2) + F_2^N(q^2)\,.
\label{eq:GEGM}
\ena 
The slopes of these form factors are related to the well-known 
electromagnetic radii of nucleons: 
\bea 
\la r^2_E \ra^N &=& 6 \frac{dG_N^E(q^2)}{dq^2}\bigg|_{q^2 = 0} \,, \\
\la r^2_M \ra^N &=& \frac{6}{G_M^N(0)} \,
\frac{dG_M^N(q^2)}{dq^2}\bigg|_{q^2 = 0}  \,. 
\ena 

We would like to emphasize that reproducing data on the neutron charge radius
$\la r^2_E \ra^n$ is a nontrivial task (see e.g. 
discussion in Ref.\cite{deAraujo:2003ke}). 
As well-known the naive nonrelativistic quark model based on SU(6)
spin-flavor symmetry implies $\la r^2_E \ra^n \equiv 0$. The dynamical
breaking of the SU(6) symmetry based on the inclusion of the quark spin-spin
interaction generates a nonvanishing value of $\la r^2_E \ra^n$.
From this point of view the
dominant contribution to the $\la r^2_E \ra^n $ comes from
the Pauli term:
\bea
\la r^2_E \ra^n \simeq \frac{6}{4m^2_{N}} F_2^n(0) \,. 
\ena

The experimental data on the nucleon
Sachs form factors in the space-like region $Q^2=-q^2\ge 0$
can be approximately described by the dipole approximation
\be
G^p_E(q^2)\approx \frac{G^p_M(q^2)}{1+\mu_p} 
\approx \frac{G^n_M(q^2)}{\mu_n}
\approx \frac{4m_N^2}{q^2}\frac{G^n_E(q^2)}{\mu_n} 
 \approx \frac{1}{\left(1-q^2/0.71\,{\rm GeV}^2\right)^2}\equiv D_N(q^2)\,.
\en 
According to present data the dipole approximation
works well up to 1 GeV$^2$ (with an accuracy of up to 25\%). For higher values
of $Q^2$ the deviation of the nucleon form factors from the dipole 
approximation becomes more pronounced.
In particular, the best description of magnetic moments, electromagnetic 
radii and form factors is achieved when we consider a superposition of the
$V$-- and $T$--currents of nucleons according to Eq.~(\ref{eq:superpo}) 
with $x = 0.8$. For the size parameter of the nucleon we take 
$\Lambda_N = 0.5$ GeV.

In Table I we present the results for the magnetic moments and 
electromagnetic radii for this set of model parameters. 
In Fig.~3 we present our results for the $q^{2}$ dependence of electromagnetic 
form factors in the 
region $Q^2\in [0,1]\,{\rm GeV}^2$. Fig.~3 also shows plots of the dipole 
approximation to the form factors. The agreement of our results with the 
dipole approximation is satisfactory. Inclusion of chiral corrections as,
for example, developed and discussed in~\cite{Faessler:2005gd} may lead to a
further improvement in the low $Q^{2}$ description.  

\section{$\Lambda$--type mass  function and electromagnetic vertex}

In a future publication we plan to study the rare baryon decays
$\Lambda_{b} \to \Lambda_{s}\,\ell^{+}\ell^{-}$ in the context of the
covariant quark model~\cite{next}. It is the purpose of this section to 
provide the necessary material that allows for a covariant quark model
description of the 
$\Lambda = (Q[ud])$-type baryons composed of a $(s,c,b)$ quark Q and a light 
diquark-like state $[ud]$ with spin and isospin zero.

In general, for the 
$\Lambda$-type baryons one can construct three types of currents 
without derivatives --- pseudoscalar $J^P$, scalar $J^S$ and 
axial-vector $J^A$ 
(see, Refs.~\cite{Faessler:2001mr,Branz:2010pq,
Ivanov:1996fj,Ivanov:1999bu}):   

\bea
J^P_{\Lambda_{Q[ud]}}&=& \epsilon^{a_1a_2a_3} \, 
Q^{a_1} \, u^{a_2} C \gamma_5 d^{a_3} \,, 
\nn
J^S_{\Lambda_{Q[ud]}}&=& \epsilon^{a_1a_2a_3} \, \gamma^5 \, 
Q^{a_1} \, u^{a_2} C  d^{a_3} \,,
\nn
J^A_{\Lambda_{Q[ud]}}&=& \epsilon^{a_1a_2a_3} \, 
\gamma^\mu \, Q^{a_1} \, u^{a_2}  
C\gamma_5\gamma_\mu d^{a_3} \,. 
\label{eq:Lambda-cur1}
\ena 
There are only two independent linear combinations 
of the above three currents given by $J^{V}=(2J^{P}-2J^{S}+J^{A})/3$ and
$J^{T}=J^{P}+J^{S}$.
The symbol $[ud]$ denotes antisymmetrization
of both flavor and spin indices w.r.t. the light quarks $u$
and $d$. We will consider three flavor types of the  
$\Lambda$-baryons: 
$\Lambda^0_s[ud]$, $\Lambda^+_c[ud]$ and $\Lambda^0_{b}[ud]$. 
In Ref.~\cite{Faessler:2009xn} we have shown 
that, in the nonrelativistic limit, the $J^P$ and $J^A$ interpolating currents
of the $\Lambda_{Q[ud]}$ baryons become degenerate and attain 
the (same) correct nonrelativistic limit  
(in the case of single-heavy baryons this limit coincides with the heavy 
quark limit), while the 
$J^S$ current vanishes in the nonrelativistic limit. 
On the other hand, the $J^P$ and $J^A$ interpolating currents of 
$\Lambda$-type baryons become degenerate with their 
SU(N$_{\mathrm f}$)-symmetric currents in the nonrelativistic limit. 
In Ref.~\cite{Ivanov:1996fj} we have shown that in case of 
the heavy-to-light baryon transition 
$\Lambda_c^+ \to \Lambda^0 e^+ \nu_e$ the use of a SU(3) symmetric current
for the $\Lambda^0$ hyperon is essential in order to describe data on 
$\Gamma(\Lambda_c^+ \to \Lambda^0 e^+ \nu_e)$ 
(see also discussion in Refs.~\cite{Korner:1994nh,Cheng:1995fe}). 
%One can show that our $J^P$ and $J^A$ currents become degenerate with 
%the SU(N$_{\mathrm f}$)-symmetric currents. 
Therefore, in the following we restrict ourselves 
to the simplest pseudoscalar $J^P$  current. 
The nonlocal interpolating  three--quark current is written as 
\bea
J_\Lambda(x) &=& \int\!\! dx_1 \!\! \int\!\! dx_2 \!\! \int\!\! dx_3 \, 
F_\Lambda(x;x_1,x_2,x_3) \, J^{(\Lambda)}_{3q}(x_1,x_2,x_3)\,,
\label{eq:Lambda-cur2}\\
J^{(\Lambda)}_{3q}(x_1,x_2,x_3) &=& 
\tfrac12 \epsilon^{a_1a_2a_3} \, Q_{a_1}(x_1)\,
\left( u^{a_2}(x_2) \,C \, \gamma^5 \, d^{a_3}(x_3)
      -d^{a_2}(x_3) \,C \, \gamma^5 \, u^{a_3}(x_2) \right) 
\nn
&=&  \epsilon^{a_1a_2a_3} \, Q^{a_1}(x_1)\,
 u^{a_2}(x_2) \,C \, \gamma^5 \, d^{a_3}(x_3)\,,
\nn
\nn
\bar J_\Lambda(x) &=& \int\!\! dx_1 \!\! \int\!\! dx_2 \!\! \int\!\! dx_3 \, 
F_\Lambda(x;x_1,x_2,x_3) \, \bar J^{(\Lambda)}_{3q}(x_1,x_2,x_3)\,,
\nn
\bar J^{(\Lambda)}_{3q}(x_1,x_2,x_3) &=& 
\epsilon^{a_1a_2a_3} \, \bar d^{a_3}(x_3)\, \gamma^5 \,C\, \bar u^{a_2}(x_2)  
\cdot \bar Q^{a_1}(x_1)\,, 
\nonumber
\ena
where $Q=s,c,b$.

The calculation of the $\Lambda$-type mass function and 
the electromagnetic vertex proceeds in the same way as 
in the nucleon case. The matrix elements in momentum space read
\be
\Sigma_\Lambda(p) = 6 g_\Lambda^2
\Bla\Bla\bar\Phi_\Lambda^2(-z_0)
S_Q(k_1+w_1 p)
\Tr\left[S_u(k_2-w_2 p)\gamma^5 S_d(k_2-k_1+w_3 p)\gamma^5\right]\Bra\Bra\,,
\label{eq:mass-lam}
\en
where we use the same short-hand notation $<<...>>$ for the two-fold 
loop-momentum integration as before (see Eqs.~(\ref{eq:short1})).
The variable $z_{0}$ is defined in~(\ref{eq:short2}).

The various contributions to the electromagnetic vertex are given by 

\bea
\Lambda^\mu_{\Lambda\, Q}(p,p')
&=& 6\, e_Q\, g_\Lambda^2
\Bla\Bla\bar\Phi_\Lambda(-z_0)
\bar\Phi_\Lambda\Big(-\tfrac12(k_1-k_2+w_3 q)^2
                   -\tfrac16(k_1+k_2+(2w_2+w_3) q)^2\Big)
\nn
&\times&
S_Q(k_1+w_1 p')\gamma^\mu S_Q(k_1+w_1 p'+q)
\Tr\left[S_u(k_2-w_2 p')\gamma^5 S_d(k_2-k_1+w_3 p')\gamma^5\right]\Bra\Bra\,,
\nn\nn
\Lambda^\mu_{\Lambda\, u}(p,p')
&=& -\,6\, e_u\, g_\Lambda^2
\Bla\Bla\bar\Phi_\Lambda(-z_0)
  \bar\Phi_\Lambda\Big(-\tfrac12(k_1-k_2+w_3 q)^2
                     -\tfrac16(k_1+k_2-(2w_1+w_3) q)^2\Big)
\nn
&\times&
S_Q(k_1+w_1 p')
\Tr\left[S_u(k_2-w_2 p'-q)\gamma^\mu S_u(k_2-w_2 p')
\gamma^5 S_d(k_2-k_1+w_3 p')\gamma^5\right]\Bra\Bra\,,
\nn\nn
\Lambda^\mu_{\Lambda\, d}(p,p')
&=& 6\, e_d\, g_\Lambda^2
\Bla\Bla\bar\Phi_\Lambda(-z_0)
  \bar\Phi_\Lambda\Big(-\tfrac12(k_1-k_2-(w_1+w_2) q)^2
                     -\tfrac16(k_1+k_2-(w_1-w_2) q)^2\Big)
\nn
&\times&
S_Q(k_1+w_1 p')
\Tr\left[S_u(k_2-w_2 p')
\gamma^5 S_d(k_2-k_1+w_3 p')\gamma^\mu S_d(k_2-k_1+w_3 p'+q) 
\gamma^5\right]\Bra\Bra\,,
\nn\nn
\Lambda^\mu_{\Lambda\,(a)}(p,p')
&=& 6\, g_\Lambda^2
\Bla\Bla\bar\Phi_\Lambda(-z_0)
\tilde E_\Lambda^\mu(k_1+w_1p',-k_2+w_2p',k_2-k_1+w_3p';q)
\nn
&\times&
S_Q(k_1+w_1 p')
\Tr\left[S_u(k_2-w_2 p')\gamma^5 S_d(k_2-k_1+w_3 p')\gamma^5\right]\Bra\Bra\,,
\nn\nn
\Lambda^\mu_{\Lambda\, (b)}(p,p')
&=& 6\, g_\Lambda^2
\la\la\bar\Phi_\Lambda(-z_0)
\tilde E_\Lambda^\mu(k_1+w_1p,-k_2+w_2p,k_2-k_1+w_3p;-q)
\nn
&\times&
S_Q(k_1+w_1 p)
\Tr\left[S_u(k_2-w_2 p)\gamma^5 S_d(k_2-k_1+w_3 p)\gamma^5\right]\Bra\Bra\,.
\label{eq:em-vertex-lam}
\ena
One now has three e.m. vertex contributions because there are three different
quarks in the $\Lambda_{Q}$ state.
The function $\tilde E_\Lambda^\mu(r_1,r_2.r_3;r)$ has been defined in 
Eq.~(\ref{eq:cur-em-nonloc}). The variables
$q_1=\sum_{i=1}^3w_{i1}r_i$ and $q_2=\sum_{i=1}^3w_{i2}r_i$
in $\tilde E_\Lambda^\mu(r_1,r_2.r_3;r)$ can be seen to be related to the
loop momenta by
\be
q_1 = \tfrac{1}{\sqrt{2}}(k_1-k_2)\,, \qquad
q_2 = -\,\tfrac{1}{\sqrt{6}}(k_1+k_2)
\label{eq:q1q2}
\en
for both bubble diagrams. By using Eq.~(\ref{eq:q1q2})
one finds the $q=0$ relations 
\bea
\Lambda^\mu_{\Lambda\,(a)}(p,p)+\Lambda^\mu_{\Lambda\,(b)}(p,p)
&=& - \,8\,g_\Lambda^2
\Bla\Bla
(Q_1 k_1^\mu + Q_2 k_2^\mu)  \bar\Phi'_\Lambda(-z_0)\bar\Phi_\Lambda(-z_0)
\nn
&\times&
S_Q(k_1+w_1 p)
\Tr\left[S_u(k_2-w_2 p)\gamma^5 S_d(k_2-k_1+w_3 p)\gamma^5\right]\Bra\Bra\,,
\nn\nn
Q_1 &=& e_1 (w_2+2w_3) - e_2(w_1-w_3)  - e_3 (2w_1+w_2)\,, 
\nn
Q_2 &=&   e_1 (w_2-w_3)  - e_2(w_1+2w_3) + e_3 (w_1+2w_2)\,.
\label{Q1Q2}
\ena
where the subscripts on the charges $e_{i}$ refer to the flavors
of the three quarks:
$"i=1"\to "s,c,b"$, $"i=2"\to "u"$ and $"i=3"\to "d"$.  
Next we will use an IBP-identity to write  
\be
\Bla\Bla\frac{\partial}{\partial k_i^\mu}
\Big\{\bar\Phi^2_\Lambda(-z_0)
S_Q(k_1+w_1 p)
\Tr\left[S_u(k_2-w_2 p)\gamma^5 S_d(k_2-k_1+w_3 p)\gamma^5\right]
\Big\}\Bra\Bra\, \equiv\, 0\,, \qquad (i=1,2)\,.
\label{eq:IP-lam}
\en
One finds
\bea
\Bla\Bla 
k_1^\mu\,A_0\Bra\Bra &=& \frac14 \Bla\Bla(2\,A^\mu_1 + A^\mu_2 - A^\mu_3 )
\Bra\Bra\,,
\nn
\Bla\Bla k_2^\mu\,A_0 \Bra\Bra &=& \frac14 
\Bla\Bla ( A^\mu_1 + 2\,A^\mu_2 + A^\mu_3 ) \Bra\Bra\,,
\label{eq:identity-lam}
\ena
where
\bea
A_0 &=& \bar\Phi'_\Lambda(-z_0)\,\bar\Phi_\Lambda(-z_0)
S_Q(k_1+w_1 p)
\Tr\left[S_u(k_2-w_2 p)\gamma^5 S_d(k_2-k_1+w_3 p)\gamma^5\right]\,,
\nn
A^\mu_1&=& \bar\Phi^2_\Lambda(-z_0)
S_Q(k_1+w_1 p)\gamma^\mu S_Q(k_1+w_1 p)
\Tr\left[S_u(k_2-w_2 p)\gamma^5 S_d(k_2-k_1+w_3 p)\gamma^5\right]\,,
\nn
A^\mu_2 &=& \bar\Phi^2_\Lambda(-z_0)
S_Q(k_1+w_1 p)
\Tr\left[S_u(k_2-w_2 p)\gamma^\mu S_u(k_2-w_2 p) 
\gamma^5 S_d(k_2-k_1+w_3 p)\gamma^5\right]\,,
\nn
A^\mu_3 &=& \bar\Phi^2_\Lambda(-z_0)
S_Q(k_1+w_1 p)
\Tr\left[S_u(k_2-w_2 p)
\gamma^5 S_d(k_2-k_1+w_3 p)\gamma^\mu S_d(k_2-k_1+w_3 p)\gamma^5\right]\,.
\label{eq:bricks}
\ena
Using these identities and collecting all pieces together, one has
\be
\Lambda^\mu_{\Lambda}(p,p)
=(e_Q+e_u+e_d)\,\frac{\partial\Sigma_\Lambda(p)}{\partial p^\mu}\,,
\qquad \not\! p=m_\Lambda\,.
\label{eq:WI-lam}
\en
As was discussed above, this Ward identity allows one to use
the compositeness condition $Z_\Lambda=0$ written in the form
\be
\Lambda^\mu_{\Lambda}(p,p) = \gamma^\mu\,,\qquad \not\! p=m_\Lambda\,,
\label{eq:Z=0-lam}
\en
where we take $e_Q=e_c$ for the present discussion.  
Again we have checked analytically that,
on the $\Lambda$-type baryon mass shell, the vertex diagrams are gauge 
invariant by themselves
and the non-gauge invariant parts coming from the bubble diagrams
corresponding to Fig.2(c) and 2(d) cancel each other before $t$-integration. 
The standard definition of the electromagnetic form factors is
\be 
\Lambda^{\mu}_{\Lambda}(p,p^{\prime}) = 
\gamma_{\mu}F_{1}(q^2)-\frac{i\sigma^{\mu q}}{2m_{\Lambda}} F_2(q^2)\,,
\label{eq:em-ff-lam}
\en 
where 
$\sigma^{\mu q}=\tfrac{i}{2}(\gamma^\mu\gamma^\nu-\gamma^\nu\gamma^\mu)q_\nu.$ 
The magnetic moment of the $\Lambda$-type baryon is defined by 
\be 
\mu_\Lambda \, = \, 
\left( \, F_{1}(0) + F_{2}(0) \, \right) \,\, \frac{e}{2 m_\Lambda} \,. 
\en 
In terms of the nuclear magneton (n.m.) 
$\frac{e}{2 m_p}$  
%($\hbar=1$) 
the   $\Lambda$-type baryon  magnetic moment 
%\mu_N = \frac{e}{2 m_p}$  ($\hbar=1$) 
the $\Lambda$--hyperon magnetic moment is given by 
\be 
\mu_\Lambda  = 
\ ( \, F_1(0) + F_2(0) \, ) \,\, \frac{m_p}{m_\Lambda}\,,  
%\qquad {\rm (in \ units \ of \ n. m.)} 
\en 
where $m_p$ is the proton mass. 

In the present paper we shall only make a rather cursory investigation
into the possible values of the size parameters of the 
$\Lambda = (Q[ud])$-type 
baryons. A more detailed investigation will be left to our future 
publication~\cite{next} where we 
will include information on the charged current transitions 
$\Lambda_{b} \to \Lambda_{c}$ and $\Lambda_{c} \to \Lambda_{s}$ to specify the
values of the size parameters of the $\Lambda = (Q[ud])$-type 
baryons. 

Let us assume for the moment that the size parameters are the
same for all $\Lambda$-type baryons. One then has 
\bea
\Lambda_\Lambda &=& 0.5\, {\rm GeV} 
\quad \mu_{\Lambda_s} = -0.73\,, 
\quad \mu_{\Lambda_c} = +0.36\,,
\quad \mu_{\Lambda_b} = -0.06\,,
\nn
\Lambda_\Lambda &=& 1.0\, {\rm GeV} 
\quad \mu_{\Lambda_s} = -0.68\,, 
\quad \mu_{\Lambda_c} = +0.40\,,
\quad \mu_{\Lambda_b} = -0.06\,,
\nn
\Lambda_\Lambda &=& 1.5\, {\rm GeV} 
\quad \mu_{\Lambda_s} = -0.61\,,
\quad \mu_{\Lambda_c} = +0.44\,,
\quad \mu_{\Lambda_b} = -0.07\,. 
\label{eq:mag-mom-lam}
\ena 
The magnetic moment of the $\Lambda_{s}$ has to be compared with the 
experimental value listed in~\cite{Nakamura:2010zzi}  
\be
\mu_{\Lambda_s} = -\,0.613 \pm 0.004 \,. 
%\ (\mathrm{in n.m.}) \,.
\label{eq:mus}
\en
%\be
%\mu_{\Lambda_s} = (-\,0.613 \pm 0.004\, )\,\mu_N\,.
%\label{eq:mus}
%\en$\Lambda = (Q[ud])$-type baryons in our future publication~\cite{next}.
Eq.~(\ref{eq:mag-mom-lam}) shows that the value of the magnetic moment
of the $\Lambda_{s}$ is quite stable against variations of its size parameter.
There is no experimental information on the magnetic moments of the
$\Lambda_{b}$ and $\Lambda_{c}$. 

The calculation of the form factors in our approach is automated by the use 
of FORM and FORTRAN packages written for this purpose. In order to be
able to compare with our earlier unconfined calculations we have written
two versions for the confined and the unconfined versions
of the covariant quark model.

%-----------------------------------------------------------------------------

\section{Summary and conclusions}

We have extended our previous formulation of the confined 
covariant quark model for mesons and tetraquark states to the baryon sector.
We have discussed in some detail various calculational aspects of the two-loop
baryon problem such as the evaluation of the baryon mass operator and its
derivative, the implementation of confinement in the two--loop context, 
the calculation of electromagnetic
current-induced transition matrix elements and the analytical verification of 
the pertinent Ward and Ward--Takahashi identities associated with the
electromagnetic matrix elements.

In our numerical work we have used the same values of the constituent quark 
masses and infrared cutoff as had been obtained before in the meson sector by a
fit to various mesonic transition matrix elements. In this way the number of
model parameters were kept to a minimum.

Using two parameters we have calculated the nucleon magnetic moments and 
charge radii as well as the electromagnetic form factors at low momentum 
transfers. An extension of our work to the
$N-\Delta(1236)$ transition can be done along the lines described in 
\cite{Faessler:2006ky}. 

We have also discussed light and heavy $\Lambda=(Q[ud])$-type baryons.
In particular we obtained a value for the size parameter of the
$\Lambda_{s}$ by a fit to its experimentally known magnetic moment.
By determining the properties of the $\Lambda=(Q[ud])$-type baryons we have 
laid the groundwork for a calculation of the rare decays of the 
$\Lambda_b$-baryon (such as $\Lambda_b \to \Lambda_s \ell^{+}\ell^{-}$) 
within the framework of the covariant quark model. 

%-----------------------------------------------------------------------------

\begin{acknowledgments}

This work was supported by the DFG under Contract No. LY 114/2-1, 
by the Federal Targeted Program ``Scientific and scientific-pedagogical 
personnel of innovative Russia'' Contract No.02.740.11.0238.
M.A.I.\ acknowledges the support of the
Forschungszentrum of the Johannes Gutenberg--Universit\"at Mainz
``Elementarkr\"afte und Mathematische Grundlagen (EMG)''
and Russian Fund of Basic Research grant No. 10-02-00368-a.

\end{acknowledgments}

\begin{table}[ht]
\begin{center}
\caption{Electromagnetic properties of nucleons~\label{tab:nuc_res}}

\vspace*{.2cm}

\def\arraystretch{1.5}
%    \begin{tabular}{|c|c|c|c|}
    \begin{tabular}{|c|c|c|}
      \hline
Quantity & Our results & Data~\cite{Nakamura:2010zzi}  \\
\hline
$\mu_p$ (in n.m.)          &  2.96       &  2.793              \\
\hline
$\mu_n$ (in n.m.)          & -1.83       & -1.913              \\
\hline
$r_E^p$ (fm)     &  0.805 &  0.8768 $\pm$ 0.0069 \\
\hline
$\la r^2_E \ra^n$ (fm$^2$) & -0.121 & -0.1161 $\pm$ 0.0022 \\
\hline
$r_M^p$ (fm)     &  0.688 &  0.777  $\pm$ 0.013 $\pm$ 0.010 \\
\hline
$r_M^n$ (fm)     &  0.685 &  0.862$^{+0.009}_{-0.008}$     \\
\hline
\end{tabular}
\end{center}
\end{table}

\vspace*{.5cm}

%\newpage 

\vspace*{1cm} 
\begin{figure}[ht]
\begin{center}
\epsfig{figure=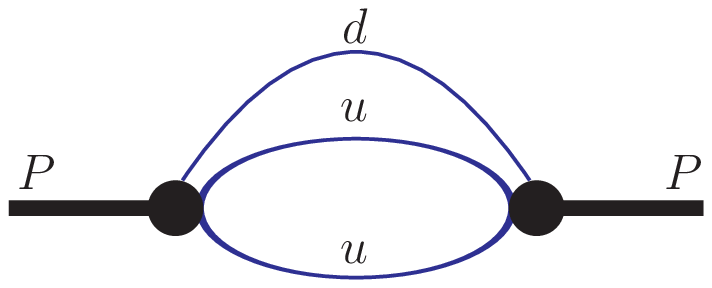,scale=.85} 

\caption{Proton mass operator.} 
\end{center}
%\end{figure} 

\vspace*{1cm} 

%\begin{figure}
\begin{center}
\epsfig{figure=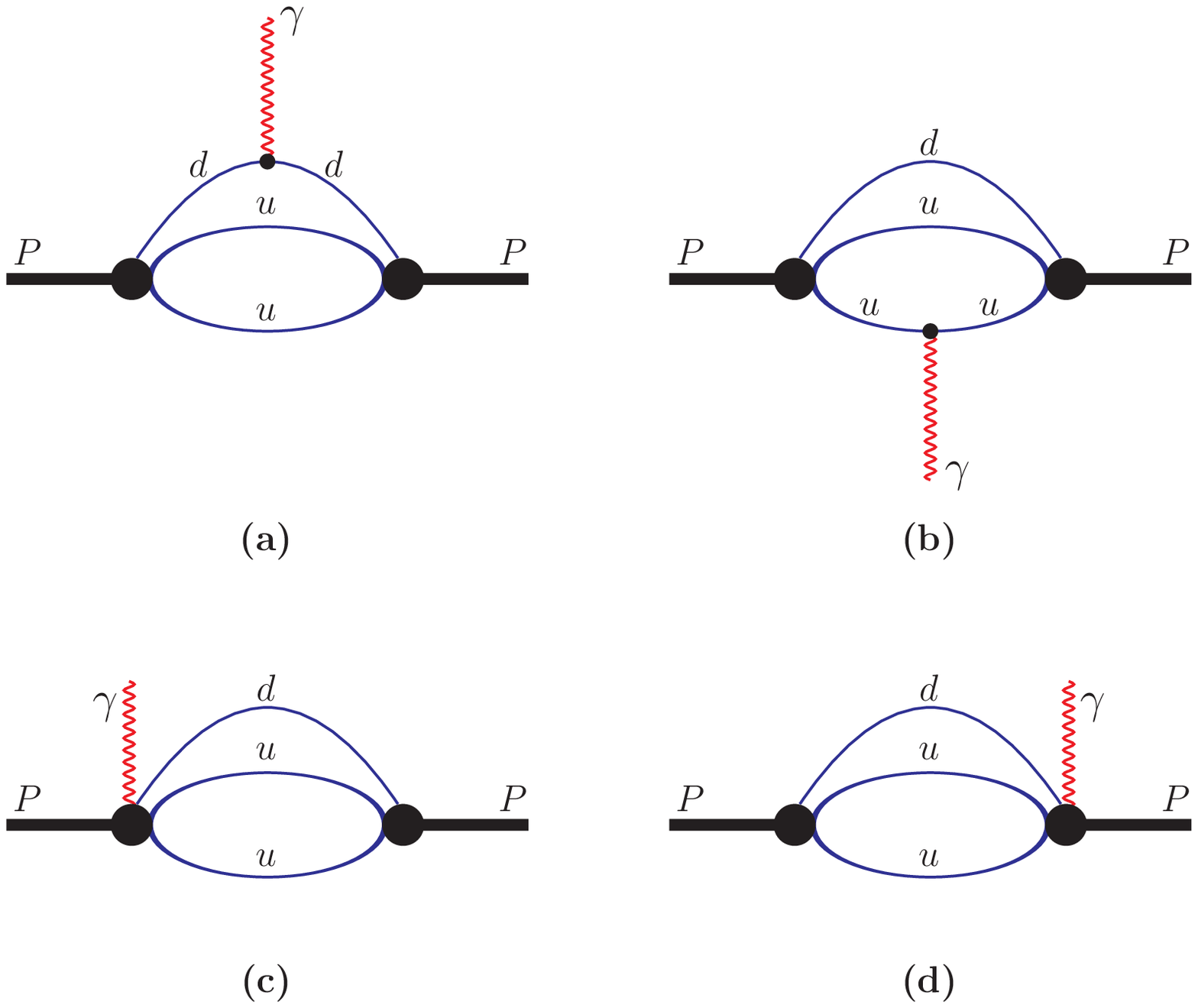,scale=.85} 
\end{center}
\caption{Electromagnetic vertex function of the proton: 
(a) vertex diagram with the e.m. current attached to d-quark; 
(b) vertex diagram with the e.m. current attached to u-quark; 
(c) bubble diagram with the e.m. current attached to the initial state vertex; 
(d) the bubble diagram with e.m. current attached to the final state vertex.}   
\end{figure} 

\begin{figure}
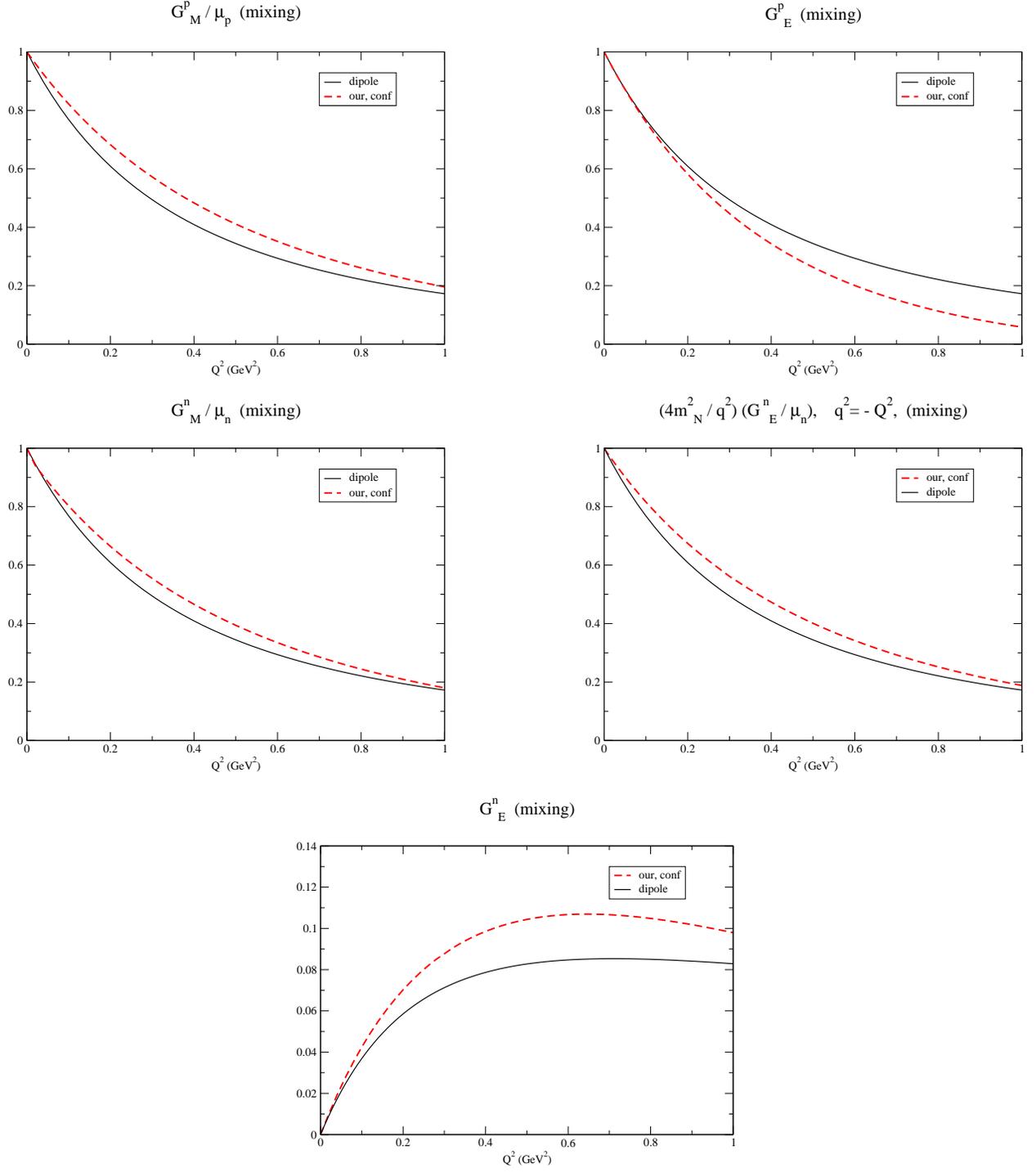

\vspace*{2cm} 
\includegraphics[width=0.40\textwidth]{GMp.conf.mix.eps} \hspace*{2cm} 
\includegraphics[width=0.40\textwidth]{GEp_N.conf.mix.eps}\\[2ex]

\includegraphics[width=0.40\textwidth]{GMn.conf.mix.eps} \hspace*{2cm} 
\includegraphics[width=0.40\textwidth]{GEn_N.conf.mix.eps}\\[2ex]

\begin{center}
\includegraphics[width=0.40\textwidth]{GEn.mix.eps}
\caption{Sachs nucleon form factors in comparions with the dipole 
representation in the space--like region $Q \le 1$ GeV$^2$.}
\end{center}
\end{figure}  

\end{document}